# Nonlinear polariton waves in dielectric medium

**Igor V. Dzedolik, Olga Karakchieva**

*Taurida National V. I. Vernadsky University, 4, Vernadsky Avenue,*
*95007, Simferopol, Ukraine*
*dzedolik@crimea.edu , karakchieva@crimea.edu*

We theoretically investigate the properties of phonon-polariton inhomogeneous harmonic wave, cnoidal wave and spatial soliton propagating in boundless dielectric medium and compute the shape of nonlinear vector polariton wave. We obtain analytically the envelopes of linearly polarized nonlinear polariton waves in the self-focusing and self-defocusing media.
*OCIS codes*: 190.3270  Kerr effect, 190.4400   Nonlinear optics, materials, 190.4223 Nonlinear wave mixing, 190.4380 Nonlinear optics, four-wave mixing,

## 1. INTRODUCTION

The properties and dynamics of the nonlinear waves in dielectric media still attract attention of researchers in connection with problems of propagation and controlling of high-power laser beams in the media [1 - 12]. The instability of wave in nonlinear media was investigated analytically and numerically in Refs. [1 – 7]. The properties of spatial vector soliton were described in Refs. [4 – 6, 8 - 11], and the cnoidal waves were investigated in the Kerr-type medium in Ref. [5] as the electromagnetic waves in nonlinear dielectric media. The existence of dynamically stable nonlinear defect kink modes at an interface separating a defocusing Kerr medium was investigated in Ref. [12].

The electromagnetic wave falling in dielectric medium propagates as bound electromagnetic and phonon waves forming the polariton wave [2, 4, 7, 13 - 16]. In this paper we theoretically investigate the properties of phonon-polariton wave propagating as inhomogeneous harmonic wave, cnoidal wave and spatial soliton in the Kerr-type medium, and analyze the conditions of the wave transformations. We have shown that polariton wave representing as polariton flow decomposes at two flows with different directions, i.e. the wavevector of polariton wave gains two values. The envelope of polariton wave at transverse plane can transform to the cnoidal wave or spatial soliton depending on the field density and medium parameters in the self-focusing medium. In the self-defocusing medium the singularity point appears at transverse plane of polariton wave.

## 2. THEORETICAL MODEL

Consider the medium with local centers of inversion, i.e. the medium with response of the third order susceptibility $c_3$. For harmonic electromagnetic field $E \sim exp(-iwt)$ we can represent the polarization vector of medium as [16]

$$\mathbf{P} = c_1 \mathbf{E}_a \, exp(-iwt) + c_{31} E_a^2 \mathbf{E}_a \, exp(-iwt), \qquad (1)$$

where $\quad c_1 = \dfrac{1}{4p}\left(\dfrac{w_e^2}{\tilde{w}_1^2} + \dfrac{w_I^2}{\tilde{W}_1^2}\right), \qquad c_{31} = -\dfrac{1}{4p}\left(\dfrac{e^2 a_{3r} w_e^2}{m^2 \left(\tilde{w}_1^2\right)^3 \left(\tilde{w}_1^2\right)^*} + \dfrac{e_{eff}^2 a_{3R} w_I^2}{m_{eff}^2 \left(\tilde{W}_1^2\right)^3 \left(\tilde{W}_1^2\right)^*}\right)$ are the linear and

nonlinear susceptibilities of medium, $\tilde{w}_1^2 = w_0^2 - w^2 - iGw$, $\tilde{W}_1^2 = W_\perp^2 - w^2 - iGw$, $w_e^2 = 4p\,e^2 N_e m^{-1}$, $w_I^2 = 4p\,e_{eff}^2 N_C m_{eff}^{-1}$ are the electron and ion plasma frequencies; $w_0^2 = q_{1r} m^{-1}$ is the electron resonance frequency, $W_\perp^2 = q_{1R} m_{eff}^{-1}$ is the resonance frequency of lattice; $a_{3r} = q_{3r} m^{-1}$, $a_{3R} = q_{3R} m_{eff}^{-1}$; $G, q_{jr}, q_{jR}$ are the phenomenological parameters depending on linear and nonlinear properties of medium. Properties of the given medium are also described by electron $w_e$ and ion $w_I$ plasma frequencies. In this case the permittivity of medium is defined by the expression $e = 1 + 4p c_1 + 4p c_{31} E_a^2$, where the both electron and ion responses at electromagnetic field are considered.

It is well known that the plane harmonic wave is unstable in nonlinear medium [5 - 7]. Instability of the plane wave leads to its modulation and further transformation to nonlinear periodic (cnoidal) wave or to spatial soliton depending of the field and medium parameters [5, 7 - 13]. We consider the process of forming of the cnoidal wave and spatial soliton from the harmonic plane wave with the frequency $w$ in nonlinear infinite dielectric medium with the third order susceptibility. We can represent the vector equation for polariton wave as

$$\nabla \times \nabla \times \mathbf{E} + c^{-2}\ddot{\mathbf{E}} = -c^{-2} 4p\, \ddot{\mathbf{P}}. \qquad (2)$$

Assuming that the electric field is polarized in plane $(x, y)$, $\mathbf{E} = \mathbf{1}_x E_x(x, y, z) + \mathbf{1}_y E_y(x, y, z)$, we obtain from Eq. (2) the following equation set

$$\frac{\partial^2 E_x}{\partial y^2} + \frac{\partial^2 E_x}{\partial z^2} - \frac{\partial^2 E_y}{\partial x \partial y} + \frac{w^2}{c^2}(1 + 4pc_1)E_x$$

$$+ \frac{4pw^2 c_{31}}{c^2}\left(|E_x|^2 + |E_y|^2\right)E_x = 0,$$

$$\frac{\partial^2 E_y}{\partial x^2} + \frac{\partial^2 E_y}{\partial z^2} - \frac{\partial^2 E_x}{\partial x \partial y} + \frac{w^2}{c^2}(1 + 4pc_1)E_y$$

$$+ \frac{4pw^2 c_{31}}{c^2}\left(|E_x|^2 + |E_y|^2\right)E_y = 0,$$

(3)

where expressions for $c_1$ and $c_{31}$ are given after Eq. (1).

## 3. WAVEVECTOR OF POLARITON WAVE

We can obtain the particular solutions of equation set (3) as the plane wave $E_{x,y} = E_{x0,y0} \, exp\left(ik_x x + ik_y y + ikz\right)$. Then we equate the determinant of set of the algebraic equations for $E_{x0}$, $E_{y0}$ to zero, $\left(c^{-2}w^2 e - k_x^2 - k^2\right)\left(c^{-2}w^2 e - k_y^2 - k^2\right) - k_x^2 k_y^2 = 0$, and obtain the nonlinear dispersion equation

$$k^4 - \left(2\frac{w^2}{c^2}e - k_\perp^2\right)k^2 + \frac{w^2}{c^2}e\left(\frac{w^2}{c^2}e - k_\perp^2\right) = 0,$$

(4)

where $e = 1 + 4pc_1 + 4pc_{31}w$ is the nonlinear permittivity, $w = E_{x0}^2 + E_{y0}^2$ is the field density, $k_\perp = \sqrt{k_x^2 + k_y^2}$ is the transverse wavevector. We evaluate the values of longitudinal wavevector by solving the dispersion Eq. (4)

$$k_+ = \frac{w}{c}(1 + 4pc_1 + 4pc_{31}w)^{1/2},$$

$$k_- = \left[\frac{w^2}{c^2}(1 + 4pc_1 + 4pc_{31}w) - k_\perp^2\right]^{1/2}.$$

(5)

The values of the wavevectors (5) depend on the field density $w$ of the polariton wave.

We can represent the polariton wave as the carrier harmonic $\sim exp(ikz)$ propagating along the axis $z$. But the wavevector (5) has two values $k_+$ and $k_-$, thus the polariton wave takes two carrier harmonics (with wavevector $k_+$ and $k_-$), and one of them (with wavevector $k_-$) propagates at the angle of axis $z$ depending on the transverse wavevector $k_\perp$.

The linear $c_1$ and nonlinear $c_{31}$ susceptibilities of medium are complex, and the wavevectors (5) are complex too, but in the transparent medium with $G = 0$ the susceptibilities $c_1$, $c_{31}$ have real values.

## 4. ENVELOPE OF VECTOR INHOMOGENEOUS POLARITON WAVE

The inhomogeneous carrier harmonic can be described as $E_{x,y} = \tilde{E}_{x,y}(x, y, z) exp(ik z)$, where $\tilde{E}_{x,y}(x, y, z)$ are the slowly varying amplitudes of two transverse electric field components, $k = k_+$ is the wavevector along the axis $z$ (Eq. (5)). If we neglect the second derivative on $z$ from the slowly varying amplitudes $\tilde{E}_{x,y}(x, y, z)$, we obtain a set of nonlinear Schrödinger equations from the equation set (3)

$$i2k\frac{\partial \tilde{E}_x}{\partial z} + \frac{\partial^2 \tilde{E}_x}{\partial y^2} - \frac{\partial^2 \tilde{E}_y}{\partial x \partial y} + a_3\left(\left|\tilde{E}_x\right|^2 + \left|\tilde{E}_y\right|^2\right)\tilde{E}_x = 0,$$

$$i2k\frac{\partial \tilde{E}_y}{\partial z} + \frac{\partial^2 \tilde{E}_y}{\partial x^2} - \frac{\partial^2 \tilde{E}_x}{\partial x \partial y} + a_3\left(\left|\tilde{E}_x\right|^2 + \left|\tilde{E}_y\right|^2\right)\tilde{E}_y = 0,$$

(6)

where $a_3 = 4p\, c^{-2} w^2 c_{31}$; $a_3 > 0$ at $c_{31} > 0$ in the self-focusing medium, and $a_3 < 0$ at $c_{31} < 0$ in the self-defocusing medium. The set of equations (6) describes the nonlinear periodic and solitary polariton waves in nonlinear medium with the third order susceptibility. In the special case at absence of the mixed derivative of cross members, the equation set (6) describes the vector spatial solitons in infinite dielectric nonlinear medium having researched in [5].

In the case when the vector polariton wave does not change the amplitude $\tilde{E}_{x,y}(x, y, z)$ along the axis $z$, we can define the dependence of field on longitudinal coordinate by the constant phase displacement $q$ as $\tilde{E}_j = e_j\, exp(iqz)$, where $e_j$ is the real amplitude, $j = x, y$. In this case we obtain from (6) the equation set for amplitude of stable inhomogeneous wave

$$\frac{\partial^2 e_x}{\partial y^2} - \frac{\partial^2 e_y}{\partial x \partial y} + a_1 e_x + a_3\left(e_x^2 + e_y^2\right)e_x = 0,$$

$$\frac{\partial^2 e_y}{\partial x^2} - \frac{\partial^2 e_x}{\partial x \partial y} + a_1 e_y + a_3\left(e_x^2 + e_y^2\right)e_y = 0,$$

(7)

where $a_1 = c^{-2}w^2(1 + 4pc_1) - k^2 - 2kq$. We can obtain numerically the solutions of equation set (7) with the boundary conditions $e_j \rightarrow 0$, $de_j / dr_j \rightarrow 0$ at $|r_j| \rightarrow \infty$. The view of envelope $e_x$ or $e_y$ of stable vector polariton wave is represented in Fig. 1.

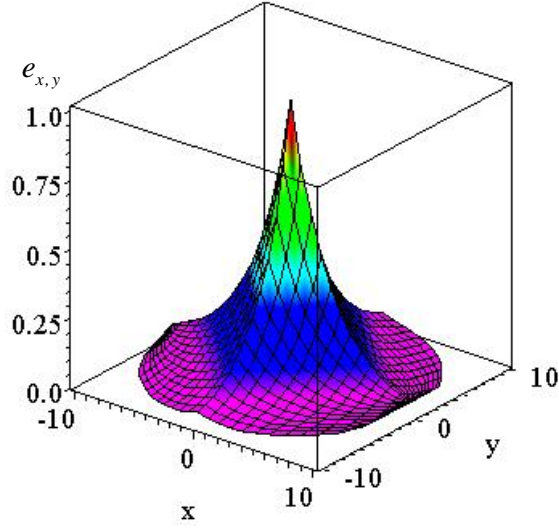

Fig. 1. The envelope of stable inhomogeneous vector polariton wave.

## 5. LINEARLY POLARIZED POLARITON WAVE

The equation set (7) is split at two uncoupled equations for linear polarization of the polariton wave, when the dispersion law $c^{-2}w^2(1 + 4pc_1) - k^2 = 0$ takes place. The equations at linear wave polarization are:

1) for the wave polarization along the axis $x$, assuming $e_y = 0$,

$$\frac{d^2 e_x}{dy^2} - a_1 e_x + a_3 e_x^3 = 0,$$ (8)

2) for the wave polarization along the axis $y$, assuming $e_x = 0$,

$$\frac{d^2 e_y}{dx^2} - a_1 e_y + a_3 e_y^3 = 0,$$ (9)

where $a_1 = 2kq$. The sign "plus" before the nonlinear term in equations (8) and (9) when $a_3 > 0$ characterizes the self-focusing medium, where the bright spatial solitons can form.

We can obtain analytically the solutions of equation (8) and (9) for bright polariton solitons with linear polarization. The first integral of the both equations (8) and (9) looks like $\left(de/dx\right)^2 = a_1 e^2 - a_3 e^4/2 + C$, where $C$ is an integration constant, $x = (x,y)$. The boundary conditions for the soliton $e \to 0$, $de/dx \to 0$ at $|x| \to \infty$ allow to define the integration constant as $C = 0$. The boundary conditions for soliton centre $e(0) = const$, $de(0)/dx = 0$, disposed in the point $x = 0$, allow to define the phase displacement as $q = a_3 e^2(0)/4k$. The second integral of the equations for bright soliton looks like $\sqrt{|a_3|/2}\,x = \int e^{-1}\left(e^2(0) - e^2\right)^{-1/2} de$, and after its integration we obtain $e(x) = e(0) \cosh^{-1}\left(e(0)\sqrt{|a_3|/2}\,x\right)$. In this case the polariton waves look like the spatial bright solitons polarized along the axis $x$ or $y$, accordingly,

$$\widetilde{E}_x(y,z) = e_x(0)\cosh^{-1}\left(\frac{e_x(0)\sqrt{|a_3|}}{\sqrt{2}}\,y\right)\exp\left(i\frac{a_3 e_x^2(0)}{4k}z\right),\ \widetilde{E}_y = 0\,; \tag{10}$$

$$\widetilde{E}_y(x,z) = e_y(0)\cosh^{-1}\left(\frac{e_y(0)\sqrt{|a_3|}}{\sqrt{2}}\,x\right)\exp\left(i\frac{a_3 e_y^2(0)}{4k}z\right),\ \widetilde{E}_x = 0\,. \tag{11}$$

Besides the spatial solitons there are the cnoidal waves as the solutions of equations (8) and (9) in the self-focusing medium with $a_3 > 0$. We choose the boundary conditions as $e = const$, $de/dx = 0$ at $|x| \to \infty$ for cnoidal wave, i.e. the integration constant is not equal zero $C = a_3 e_\infty^4/2 - a_1 e_\infty^2 = const$. Then the second integral looks like $\sqrt{|a_3|/2}\,x = \int_0^e \left(C' + a'e^2 - e^4\right)^{-1/2} de$, where $a' = 2a_1 a_3^{-1}$, $C' = 2Ca_3^{-1}$, and phase displacement is equal $q = a_3 e_\infty^2/4k - C/2k e_\infty^2$. In the self-focusing medium we obtain the envelopes of cnoidal polariton waves with polarization along the axis $x$ or axis $y$ in the form of elliptic cosine,

$$\widetilde{E}_x(y,z) = \widetilde{e}_x\, cn\left(K_x y, \widetilde{k}_x\right)\exp(iq_1 z),\ \widetilde{E}_y = 0\,; \tag{12}$$

$$\widetilde{E}_y(x,z) = \widetilde{e}_y\, cn\left(K_y x, \widetilde{k}_y\right)\exp(iq_2 z),\ \widetilde{E}_x = 0\,, \tag{13}$$

where $\widetilde{e}_j = \left[a'/2 + \left(a'^2/4 + C_j'\right)^{1/2}\right]^{1/2}$, $a_j = \left(a'^2/4 + C_j'\right)^{1/4}\sqrt{|a_3|}$, $q_{1,2} = a_3 e_{\infty(x,y)}^2/4k - C/2k e_{\infty(x,y)}^2$, $\widetilde{k}_j = \left[2 + a'\left(a'^2/4 + C_j'\right)^{-1/2}\right]^{1/2}/2$ is the modulus of elliptic integral.

The cnoidal wave can transform to the spatial soliton in the special case [7, 9, 12, 14, 16]. The elliptic cosine is transformed to the hyperbolic secant $cn(x,1) \rightarrow cosh^{-1}(x)$ describing the spatial soliton at the modulus $\tilde{k} \rightarrow 1$, when $C \rightarrow 0$, i.e. $e_\infty = 0$.

In the self-defocusing medium at $a_3 < 0$ we obtain the solutions of equations (8) or (9) in the form of elliptic tangent

$$e(x) = \tilde{e}\, tn\left(Kx, \tilde{k}_{-1}\right), \tag{14}$$

where $\tilde{e} = \left[a'/2 - \left(a'^2/4 - C'\right)^{1/2}\right]^{1/2}$, $K = \left[-a'/2 + \left(a'^2/4 - C'\right)^{1/2}\right]^{1/2} \sqrt{|a_3|/2}$,

$\tilde{k}_{-1} = \left[2\left(a'^2/4 - C'\right)^{1/2} - a'\right]^{1/2}\left[-a'/2 + \left(a'^2/4 - C'\right)^{1/2}\right]^{-1/2}$. The solution (14) demonstrates how the plane polariton wave gains the singular properties at transverse plane in the self-defocusing medium, because the elliptic tangent has the singular points.

## 6. CONCLUSION

The plane polariton wave is unstable in nonlinear medium, its transverse envelope can transform to the cnoidal wave or spatial soliton in the self-focusing medium. The transverse envelope of nonlinear vector polariton wave looks like a cone with curvilinear tangent. The forms of transverse envelopes of linearly polarized polariton waves can be the hyperbolic secant (the soliton) or elliptic cosine (the cnoidal wave) in the self-focusing medium. In the first case the polaritons propagate as the single flow, in the second case the polariton wave decomposes at the several flows. In the self-defocusing medium the transverse envelope of polariton wave gets the singular points.

Thus, the form of transverse envelope of polariton wave depends on the wave intensity and parameters of medium and it can be modified by changing the intensity or medium parameters. If the polariton wave is radiated from the medium into air, we can control the form of the wave in this way.

### ACKNOWLEDGMENT

The authors are grateful to Yuri S. Kivshar and Anton S. Desyatnikov for discussions about the paper.

**REFERENCES**

1. Bloembergen N., *Nonlinear optics* (New-York – Amsterdam, W.A.Benjamin, Inc., 1965).

2. Klyshko D. N., *Quantum and nonlinear optics* (Moscow: Nauka, 1980) (in Russian).

3. Sukhorukov A.P., *Nonlinear wave interactions in optics and radiophysics*, (Moscow: Nauka, 1988) (in Russian).

4. Shen Y. R. *The principles of nonlinear optics* (New York: John Wiley & Sons, 1984).

5. Kivshar Yu. S,, Agraval G. P. *Optical solitons: from fibers to photonic crystals* (New York: Academic Press, 2003).

6. Boyd R.W., Nonlinear optics, (San Diego: Academic Press, 2003).

7. Dzedolik I. V. *Polaritons in optical fibers and dielectric resonators* (Simferopol: DIP, 2007) (in Russian).

8. Xu J., Shandarov V., Wesner M., Kip D., "Observation of two-dimensional spatial solitons in iron-doped barium-calcium titanate crystals", Phys. Stat. Sol. A, **189**, No. 1, R4-R5 (2002).

9. Dzedolik I. V., "Transformation of sinusoidal electromagnetic and polarization waves into cnoidal waves in an optical fibre", *Ukr. J. Phys. Opt.*, **9**, No. 4., P. 226–235 (2008).

10. Ouyang S., Guo Q., "Dark an gray spatial optical solitons in Kerr-type nonlocal media", Opt. Exp., **17**, No. 7, P. 5170-5175 (2009).

11. Buccoliero D., Desyatnikov A.S., "Quasi-periodic transformations of nonlocal spatial solitons", Opt. Exp., **17**, No. 12, P. 9608-9613 (2009).

12. Zhong S, Huang C., Li C., Dong L., "Surface defect kink solitons", *Opt. Com.*, **285**, Is.17, P. 3674–3678 (2012).

13. Dzedolik I. V., "Period variation of polariton waves in optical fiber", *J. Opt. A: Pure Appl. Opt.* **11** 094012 (2009).

14. Dzedolik I. V., Lapayeva S. N., "Mass of polariton in different dielectric media ", *J. Opt.*, **13**, 015204 (2011).

15. Dzedolik I. V., Karakchieva O. S. , "Polaritons in nonlinear medium: generation, propagation and interaction", *Proc. NLP\*2011, IEEE* Catalog Number: CFP1112P-CDR ISBN: 978-1-4577-0479-6.

16. Dzedolik I.V., Karakchieva O., "Polariton waves in nonlinear dielectric medium", *arXiv:1205.1301*, 15 p. (2012).